\documentclass[a4paper,french]{rnti}

\usepackage[T1]{fontenc}
\usepackage[latin1]{inputenc}
\usepackage{subfigure}
\usepackage{url}
\usepackage{acronym}
\usepackage[procnumbered,ruled]{algorithm2e}
\usepackage{etoolbox}

\makeatletter
\AtBeginEnvironment{function}{\let\c@algocf\c@function}
\makeatother
\usepackage[noend]{algpseudocode}
\usepackage{algcompatible}
\algblockdefx{FORALLP}{ENDFAP}[1]%
  {\textbf{For}#1 \textbf{Do in Parallel}}%
  {\textbf{End For}}
\algloopdefx{RETURN}[1][]{\textbf{Return} #1}
\usepackage{graphicx}
\usepackage{latexsym}
\usepackage{float}
\usepackage{epsfig}
\usepackage{array}

\usepackage{fancyhdr}

\titrecourt{Classification non supervisée des données hétérogènes à large échelle}

\nomcourt{M.A. Zoghlami et al.}

\titre{Classification non supervisée des données hétérogènes à large échelle}

\auteur{Mohamed Ali Zoghlami\affil{1},
        Olfa Arfaoui\affil{2}, Minyar Sassi Hidri\affil{3}, Rahma Ben Ayed\affil{4}}

\affiliation{
    $^{*,**,***,****}$Université de Tunis El Manar\\
      \'Ecole Nationale d'Ingénieurs de Tunis, BP. 37, Le Belvédère 1002, Tunis, Tunisia\\
          $\{$$^*$medali.zoghlami,$^{**}$olfa.arfaoui,$^{***}$minyar.sassi,$^{****}$rahma.benayed$\}$@enit.rnu.tn\\
          }

\resume{%
Quand il sera question de classifier des données massives, le temps de réponse, l'accès disque et la qualité des classes formées deviennent des enjeux majeurs pour les entreprises. C'est dans ce cadre que nous avons été amenés à définir un cadre de classification non supervisée des données hétérogènes à large échelle qui contribue à la résolution de ces enjeux. Le cadre proposé s'articule autour de l'analyse descriptive à base d'Analyse des Correspondances Multiples (ACM) d'une part, et d'autre part du paradigme de programmation parallèle MapReduce dans un environnement à large échelle. Les résultats obtenus sont encourageants et prouvent l'efficience du déploiement hybride sur le volet qualité et temps de réponse autant sur des données qualitatives et quantitatives.
}

\summary{%
When it comes to cluster massive data, response time, disk access and quality of formed classes becoming major issues for companies. It is in this context that we have come to define a clustering framework for large scale heterogeneous data that contributes to the resolution of these issues. The proposed framework is based on, firstly, the descriptive analysis based on MCA, and secondly, the MapReduce paradigm in a large scale environment. The results are encouraging and prove the efficiency of the hybrid deployment on response quality and time component as on qualitative and quantitative data.
}

\begin{document}

%
\section{Introduction}
Le mouvement Big Data s'accompagne du développement d'applications à visée analytique qui traitent et analysent les données pour en tirer du sens. Ces analyses sont appelées \textit{Big Analytics} ou \textit{broyage de données}. Elles portent sur des données quantitatives complexes avec des méthodes de calcul distribué. Une fois les données stockées, ou leur flux organisé, leur valorisation nécessite une phase d'analyse (Analytics). L'objectif étant de tenter de pénétrer le grand volume de données pour en comprendre les enjeux et aider à la prise de décision. Nous considérons l'utilisation, si possible au mieux, des ressources logicielles (langages, librairies, etc.) open source existantes tout en essayant de minimiser les temps de calcul et également de maîtriser les coûts humains de développement pour réaliser un éventuel prototype d'analyse.

C'est dans ce cadre  que se situe ce travail qui vise à proposer et évaluer une méthode de classification non supervisée (ou clustering) des données hétérogènes dans un environnement à large échelle à base d'Analyse de Correspondances Multiples (ACM) \citep{escofier1988analyse} tout  en utilisant le patron d'architecture MapReduce sur des serveurs dissiminés sur le Cloud.

Ce papier est organisé comme suit : la section 2 décrit le déploiement hybride du clustering des données hétérogènes dans un environnement à large échelle. Ainsi, nous détaillons également les algorithmes que nous avons développés dans la section 3. Nous discutons les résultats du classificateur proposé dans la section 4. Enfin, la section 5 conclut le papier et ouvre sur des perspectives futures.

\section{Déploiement hybride du clustering à large échelle}
\begin{figure}[t]
\begin{center}
\includegraphics[width=6.5cm]{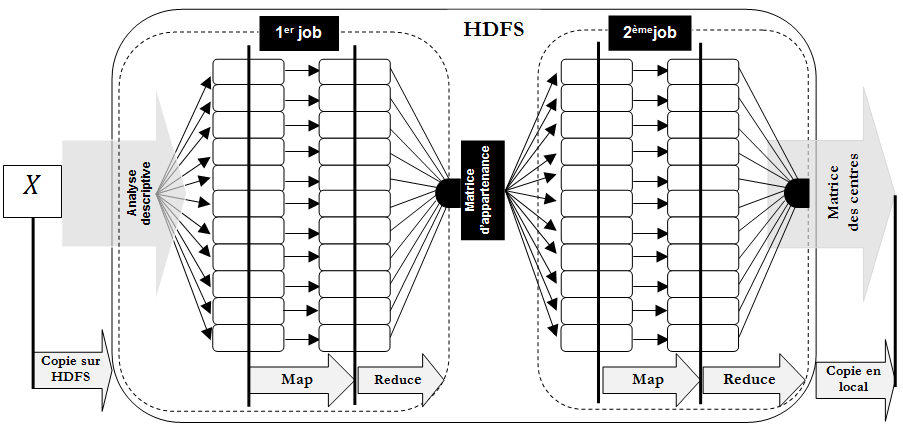}
\caption{Déploiement hybride du clustering à large échelle.}\label{Figure2.1}
\end{center}
\end{figure}

La figure \ref{Figure2.1} présente une vue globale de notre classificateur de données hétérogènes à large échelle. Les données massives correspondent à l'unique ressource qui alimente le patron de programmation MapReduce. Ainsi avant qu'elles soient soumises à une classification automatique non supervisée, les données subissent une analyse descriptive basée sur l'ACM \citep{benzecri1977histoire} pour palier le problème d'hétérogénéité.

Comme la figure \ref{Figure2.1}, les deux jobs sont complémentaires du fait que notre classificateur se base essentiellement sur l'application itérative de l'algorithme de C-Moyennes Floues (CMF) de \citet{BezdekJC1983}. Le premier job débute par une phase de pré-traitement basée sur l'analyse descriptive pour calculer et produire la matrice de partition qui sera considérée comme entrée du deuxième job et ce, afin de fournir le nombre optimal de clusters.

Pour palier le problème d'hétérogénéité et de volumétrie de données, le framework distribué RHadoop (combinaison du langage R et du framework Hadoop) a été utilisé. Il comporte deux composantes essentielles : le système de fichiers distribué Haddoop (HDFS : Hadoop Distributed File System) et le paradigme de programmation parallèle MapReduce. Ce dernier se compose de deux fonctions $Map()$ et $Reduce()$. Dans la fonction $Map()$, le n{\oe}ud analyse le problème, le découpe en sous-problèmes, et les délègue à d'autres n{\oe}uds (qui peuvent aussi en faire de même d'une manière récursive) \citep{taylor2010overview}. \`A chaque sous-problème est associée une clé qui permet de le repérer. Les sous-problèmes sont traités à l'aide de la fonction $Reduce()$ \citep{taylor2010overview}. Elle renvoie les résultats en leur associant également une clé. Le couple $<cl\acute{e}, valeur >$ tient donc une place extrêmement importante dans le dispositif MapReduce, $valeur$ étant relative, selon le cas, à des données ou à des résultats de calculs.

\section{RHClus : Clustering à base de RHadoop et MapReduce}
\begin{figure}[ht!]
\centering
    \subfigure[1$^{er}$ Job]{\label{first}
    \includegraphics[height=2.5cm]{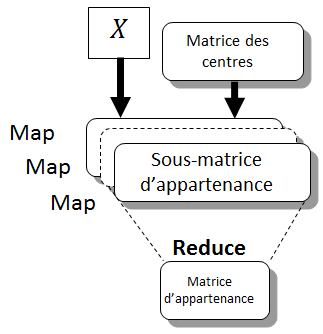}}
    \subfigure[2$^{\grave{e}me}$ Job]{\label{second}
    \includegraphics[height=2.5cm]{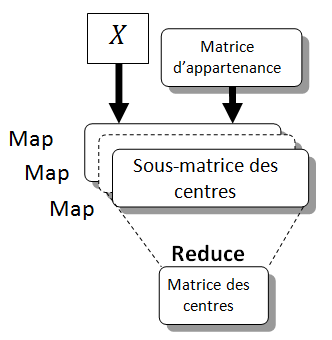}}
\caption{Jobs MapReduce.}
\label{Fd1}
\end{figure}

Nous décrivons dans cette section les deux jobs MapReduce qui présentent le socle de notre classificateur. En effet, le premier job présenté dans la figure \ref{Fd1}(a), prend comme entrée les données fusionnées horizontalement avec les $c$ centres de clusters choisis aléatoirement pour calculer et produire la matrice de partition et ce, après avoir effectué une analyse descriptive sur ces données.

La matrice de partition produite à la fin du premier job sera l'entrée du deuxième job présenté dans la figure \ref{Fd1}(b) pour produire la matrice des centres de clusters. La fonction 1 décrit les différentes étapes du mapper du premier job $MapReduce$.

\begin{function}[!htp]
 \scriptsize{
 \caption{Map du premier Job MapReduce()}\label{alg2}      
\SetAlgoLined\DontPrintSemicolon
  \SetKwFunction{algo}{algo}\SetKwFunction{proc}{Map}          
  \SetKwProg{myproc}{Fonction}{}{Fin}
  \myproc{\proc{$cl\acute{e}, valeur$}}{                                    
\lPourCh{Clé}{
  \nl Appliquer l'ACM pour transformer les relations entre les données d'entrée en des points dans un espace euclidien de faible dimension\;
  \nl Calculer les éléments de la matrice d'appartenance pour chaque clé et les stocker dans $valeur^{\prime}$\;
  \nl Émettre le couple $<cl\acute{e}^{\prime},valeur^{\prime}>$, où $cl\acute{e}$ est la liste des valeurs des enregistrements et $valeur^{\prime}$ est la matrice d'appartenance intermédiaire\;
   \textbf{Fin pour}\;}\vspace{-0.3cm}
  \nl \textbf{Retourner} ($liste(cl\acute{e}^{\prime},valeur^{\prime})$)\;}                                                 
}\end{function}

Au cours de l'exécution de la fonction $Map()$ (fonction 1), les valeurs des sous-matrices de partitions produites sont calculées en utilisant les conditions de convergence de \citep{BezdekJC1983}, et sont ensuite émises à la fonction $Reduce()$ (fonction 2).

\begin{function}[!htp]
 \scriptsize{
 \caption{Reduce du premier job MapReduce()}\label{alg3}      
\SetAlgoLined\DontPrintSemicolon
  \SetKwFunction{algo}{algo}\SetKwFunction{proc}{Reduce}          
  \SetKwProg{myproc}{Fonction}{}{Fin}
  \myproc{\proc{$cl\acute{e}, valeur$}}{                                    
\lPourCh{Clé}{
  \nl Fusionner Les sous-matrices intermédiaires  des degrés d'appartenance et stocker le résultat dans $valeur^{\prime}$\;
  \nl \'Emettre le couple $<cl\acute{e}^{\prime},valeur^{\prime}>$\;
   \textbf{Fin pour}\;}\vspace{-0.3cm}
  \nl \textbf{Retourner} ($value^{\prime}$)\;}                                                 
  }
\end{function}

La fonction 3 décrit les différentes étapes du mapper du deuxième job $MapReduce$ présenté dans le schéma \ref{Fd1} (b). Cette fonction prend comme entrée les données et la matrice de partition issues du premier job, puis elle calcule l'ensemble des sous-matrices intermédiaires des centres de clusters. Les résultats seront émis pour la fonction $Reduce()$ (fonction 4) du même job.

\begin{function}[!htp]
 \scriptsize{
 \caption{Map du deuxième Job MapReduce()}\label{alg4}      
\SetAlgoLined\DontPrintSemicolon
  \SetKwFunction{algo}{algo}\SetKwFunction{proc}{Map}          
  \SetKwProg{myproc}{Fonction}{}{Fin}
  \myproc{\proc{$cl\acute{e}, valeur$}}{                                    
\lPourCh{clé}{
  \nl Calculer les éléments de la matrice des prototypes pour chaque clé et les stocker dans $value^{\prime}$\;
  \nl Émettre le couple $<cl\acute{e}^{\prime},valeur^{\prime}>$, où $cl\acute{e}$ est la liste des valeurs des enregistrements et $valeur^{\prime}$ est l'ensemble des matrices des centres intermédiaires\;
  \textbf{Fin pour}\;}\vspace{-0.3cm}
  \nl \textbf{Retourner} ($List(cl\acute{e}^{\prime},valeur^{\prime})$)\;}                                                 
}\end{function}

La fonction 4 prend comme entrée les couples $<$clé',valeur'$>$ issus de la fonction $Map()$. Elle calcule la somme de toutes les sous-matrices des centres de clusters. Les résultats seront émis pour être copiés du HDFS vers le SGF (système de gestion de fichiers) locale.

\begin{function}[!htp]
 \scriptsize{
 \caption{Reduce du deuxième Job MapReduce()}\label{alg5}
\SetAlgoLined\DontPrintSemicolon
  \SetKwFunction{algo}{algo}\SetKwFunction{proc}{Reduce}
  \SetKwProg{myproc}{Fonction}{}{Fin}
  \myproc{\proc{$cl\acute{e}, valeur$}}{                                    
\lPourCh{clé}{
  \nl Calculer la matrice des centres en sommant les éléments des sous-matrices intermédiaires des centres et stocker le résultat dans $valeur^{\prime}$\;
  \nl \'Emettre le couple $<cl\acute{e}^{\prime},valeur^{\prime}>$\;
   \textbf{Fin pour}\;}\vspace{-0.3cm}
  \nl \textbf{Retourner} ($valeur^{\prime}$)\;}
}
\end{function}
\section{\'Evaluation de la performance}
Les expériences ont été réalisées sur la plate-forme expérimentale HP BladeSystem. Pour notre série d'expériences, nous avons utilisé le cluster vCenter hébergé par le centre de données Cloud de SESAME\footnote{http://www.universitesesame.com/campus-high-tech-sesame/data-center/}. Le déploiement a été effectué avec 20 n\oe uds. Chaque n\oe ud a un processeur Intel(R) Xeon(R) CPU E5649 \@ 2.53GHz 64 bits Red Hat CentOS version 6.5 (version du noyau 2.6.32) avec un disque dur de 100Go SCSI et 32Go de RAM.

Pour évaluer le classificateur proposé, nous avons eu recours à des jeux de données qualitatives et quantitatives issus du UCI Machine Learning Repository. Le tableau \ref{tab123} présente un résumé des informations concernant les jeux des données utilisées.

\begin{table}[ht!]
\centering
\tiny{
   \tabcolsep=1\tabcolsep
\begin{tabular}{llll}
   \hline
   Jeu de donn\'ees &\#Attributs &\#Clusters &\#Instances\\
   \hline
   Mm & 6& 2& 961   \\
   BS & 4 & 3 & 625    \\
      Taste & 9&3 & 1253   \\
   Forest&54&7&581,012\\
   Wave&21&3&	2,000,000\\
      \hline
   \end{tabular}
   }
   \caption{Détails des jeux de données.}\label{tab123}
\end{table}

\begin{figure}[ht!]
\centering
    \subfigure[Masse mammographique]{\label{first}
    \includegraphics[height=2.7cm]{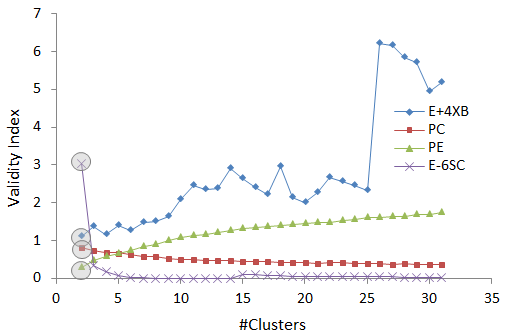}}
    \subfigure[Balance Scale]{\label{second}
    \includegraphics[height=2.7cm]{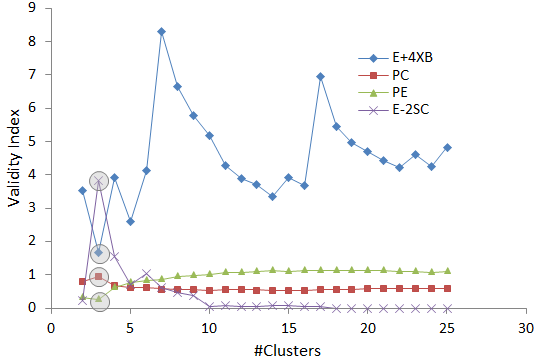}}
      \subfigure[Taste]{\label{second}
    \includegraphics[height=2.7cm]{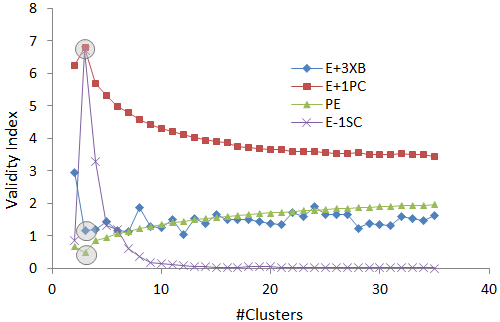}}
\caption{Indices de validité versus nombre de clusters.}
\label{FigMm}
\end{figure}

La qualité sera évaluée en fonction de la capacité à détecter le nombre optimal de clusters pour chaque jeu de données. Une série d'indices de validité a été utilisée. Nous citons essentiellement le coefficient de partition et l'entropie moyenne de \citet{Bezdek19987}, Xie-Béni de \citet{XieB91} et Separation-Compacité de \citet{dexaw}. La figure \ref{FigMm} illustre l'évolution des indices de validité par rapport aux nombre de clusters.

Selon la figure \ref{FigMm}, le classificateur proposé retourne le nombre optimal de clusters sur les différents jeux de données. L'évolution des indices de validité PC, PE, XB et SC a la même apparence et montre bien la cohérence et la stabilité du déploiement hybride sur le volet qualité.

Dans le but d'évaluer la performance du classificateur proposé, les deux jeux de données Forest et Wave ont été divisés en plusieurs sous-ensembles, comme indiqué dans le tableau \ref{tabbb}. Nous procédons à une analyse de l'évolutivité dans laquelle les tailles des ces portions sont augmentées, à chaque fois, de 100.000 (resp. 200.000) lignes pour Forest (resp. Wave) jusqu'à ce qu'on atteint la totalité de chaque jeu de données. Du fait que le nombre d'instances du jeu de données Forest est 581.012 et pour le besoin de l'expérimentation, certaines lignes ont été dupliquées pour atteindre les 600.000 instances.

\begin{table}[ht]
\centering
      \tiny{
   \begin{tabular}{l|cc|cc}
   \hline
      & \multicolumn{2}{c|}{Forest}&\multicolumn{2}{c}{Wave} \\
  \hline
  \#Instances&	Taille en octets&	Pourcentage&	Taille en octets&Pourcentage\\
   \hline
1.0E$+$05&	12839748&16.67\%&-&-\\
2.0E$+$05&25650350&33.30\%&21513887&10.00\%\\
3.0E$+$05&38466862&49.95\%&-&-\\
4.0E$+$05&51311749&66.62\%&	43026278&	20.00\%\\
5.0E+05&	64172437&	83.32\%&	-	&-\\
6.0E+05&	77017667	&100.00\%&	64537420&	30.00\%\\
7.0E+05&	-&-&		-&	-\\
8.0E+05&-&-&			86050837&	40.00\%\\
1.0E+06&-&-&			107564071&	50.00\%\\
1.2E+06&-&-&			129075962&	60.00\%\\
1.4E+06&-&-&			150588044&	70.00\%\\
1.6E+06&-&-&			172099183&	80.00\%\\
1.8E+06&-&-&			193611122&	90.00\%\\
2.0E+06&-&-&			215122859&	100.00\%\\
      \hline
   \end{tabular}
   }
\caption{Tailles et pourcentages des instances chargées au cours du déploiement.} \label{tabbb}
 \end{table}

\begin{figure}[ht!]
\centering
    \subfigure[Forest]{\label{first}
    \includegraphics[height=2.9cm]{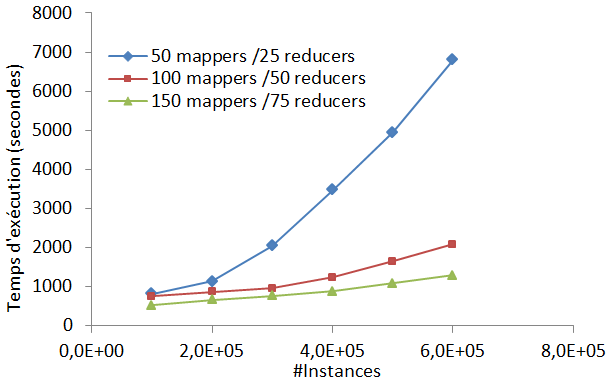}}
    \subfigure[Wave]{\label{second}
    \includegraphics[height=2.9cm]{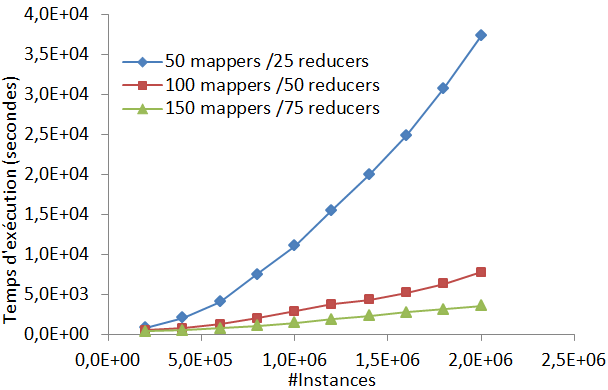}}
     \caption{Temps d'exécution versus \#instances par déploiement.}
\label{Figgg}
\end{figure}

La figure \ref{Figgg} illustre l'évolution du temps d'exécution du classificateur en fonction de la taille de la portion des données, et ce en utilisant respectivement 50 mappers et 25 reducers, 100 mappers et 50 reducers et enfin 150 mappers et 75 reducers.

En se basant sur l'allure des courbes obtenues, nous constatons qu'une baisse du nombre de mappers utilisés génère une augmentation exponentielle du temps d'exécution, alors qu'à force d'augmenter le nombre de mappers, l'allure s'aplatie et l'augmentation devient presque linéaire. Ceci montre que le classificateur proposé est vraiment adapté dans le cadre des données massives (ou Big data) dans un environnement hétérogène à large échelle.

\section{Conclusion}
Ce papier présente notre contribution dans la fouille de données hétérogènes à large échelle et plus précisément la classification non supervisée. Nous avons proposé un cadre de classification de données à base d'analyse descriptive et du paradigme MapReduce. Deux jobs ont été proposés afin de séparer les matrices des prototypes et d'appartenance.

Les résultats des expérimentations réalisées prouvent l'efficacité du système de classification proposé tenant compte de la volumétrie, de la variété et de la véracité.

Les orientations futures de ce travail concernent essentiellement l'exploitation du paradigme MapReduce dans d'autres méthodes de classification en l'occurrence les méthodes hiérarchiques et spectrales.

\bibliographystyle{rnti}
\bibliography{biblio_exemple}

\appendix


\Fr

\end{document}